\documentclass[twocolumn,12pt]{revtex4} \usepackage[dvips]{graphics,color} \def\bea{\begin{eqnarray}}
 \def\eea{\end{eqnarray}}
\def\be{\begin{equation}}
\def\ee{\end{equation}}
\def\nn{\nonumber}

\def\a{\alpha}
\def\b{\beta}

\def\e{\epsilon}

\def\r{\rho}

\def\th{\theta}

\begin{document}
\preprint{ULB-03-36}
\title{Counting the Apparent Horizon}

\vspace{1cm}
\author{\sc Arundhati Dasgupta} \email{ adasgupt@ulb.ac.be,dasgupta@aei.mpg.de}
\affiliation{Physique Theorique et Mathematique,
Universite Libre de Bruxelles,
Brussels, B-1050.}

\vspace{2cm}

\begin{abstract} Using a discrete spectrum proposed for expectation values of canonical
variables in black hole coherent states, the semiclassical entropy associated with the
Schwarzschild space-time is derived to be the area of the apparent horizon. 
 

\end{abstract}

\maketitle
The origin of black hole entropy as well as Hawking radiation is semi-classical,
and the resolution should come from a non-perturbative theory
of quantum gravity with a suitable classical limit. There have been several attempts within the realm of semi-classical physics to associate 
 discrete spectrums to the horizon area \cite{spect}. A degeneracy counting or a emission spectrum is then obtained. A more
`quantum' approach uses the area spectrum derived in loop quantum gravity 
associated with the Horizon 2-sphere which satisfies a special boundary
condition \cite{cant}. Many questions remain unanswered in the above; a couple of them are:\\
1)Why is Black Hole entropy associated with surface degrees of freedom? Or in other words why is there a principle of Holography?\\
2)Where do the degrees of freedom live: on the Horizon/ inside or outside the black hole?\\
 In this letter, we derive a degeneracy associated with the area of the
Horizon 2-sphere on a particular time slice of the Schwarzschild Black Hole
(Hence Apparent Horizon) and show, that it is indeed the {\it Entropy} of the black hole. The computation is done in Gauge Coherent States defined by Hall \cite{hall,hall2}
and applied to Canonical Gravity. As we shall see, since in these coherent states, the
classical limit of the operators can be identified, one can answer at least the two
questions stated above quite clearly. Moreover, the semiclassical spectrum associated
with Area is {\it equispaced}. Also, a concrete realisation is provided through the coherent states and the canonical operators of many of the ideas and conjectures used in previous papers regarding the `Black Hole State' and corresponding
measurements on the state.

The canonical pair $h_e(A), P^I_e(A,E)$ defined by Thiemann in \cite{tt}
is constituted by the holonomy of the SU(2) Sen-Ashtekar-Barbero-Immirzi connection $A_I^a$,
along the edges $e$ of a graph, and a momentum which is a function of the densitised triads $E^a_I$
defined on the corresponding dual graph (I is the SU(2) internal index and $a=1,2,3$ the spatial directions on a spatial slice). The coherent states are defined as functions of this canonical pair, \cite{tt,tow,th}, and constructed to be peaked at the classical values of these variables. In \cite{adg}, these states were defined
for the non-rotating black hole, and observations on the nature of the variables led to the proposal that the momentum $P^I_e(A)$ has a discrete spectrum which manifests itself when the classical value of the above variable is very small. In this letter, we show that the area operator is precisely given by the gauge invariant momentum $ P_e=\sqrt{P^I_e P^I_e}$, when the edge $e$ is radial, and has a
 equispaced spectrum. The degeneracy arises when one averages over the
microscopic quantum numbers, which are not seen by the classical expectation values.
The coherent state is a product over the coherent state for each edge of the graph, which is embedded in the classical space-time. \be
\Psi= \prod_e \psi^t_e\ee
Based on the classical value at which the state is peaked, one can determine the location of the edges as measured by the
classical metric.
Thus one can split the product over edges which lie outside the black hole, those which cross the black hole and those which lie inside the black hole. One can easily write it as a product of $\Psi^{\rm 0\equiv outside}\Psi^{\rm H\equiv horizon}\Psi^{\rm I\equiv inside}$. 
To an asymptotic observer, the inside of a black hole
remains inaccessible, and one has to trace over the coherent state which is `inside'
the black hole. This is the reason the degeneracy associated with the operators manifests itself as the `Entropy' of the black hole. Since all operators are expressed in terms of the canonical pair, one needs to study 
only the expectation values of these operators. As we observe in this letter, the `entropy' is the logarithm of the degeneracy associated with the operator $P_{e_r}$ at the {\it horizon} ($e_r$ stands for a radial edge). As per the philosophy of \cite{tt}, expectation values of {\it all other} operators must be evaluated in terms of the expectation values
of the canonical pair, and hence this degeneracy {\it cannot be broken} by any other
measurements on the black hole coherent state.

Since the area operator is associated with the momentum $P_e$, we examine it first, with the coherent state in the
momentum representation.
We then discuss the origin of the degeneracy. We count the entropy, and finally
conclude why it is the entropy of the black hole with special emphasis on the Holographic nature of the
degeneracy counting. Many of the statements here
confirm quite a few conjectures and observations made by previous investigators
of black hole entropy \cite{spect,cant}. Also, a note of caution 
must be added that these are the simplest coherent states, and all the results pertain to the
situation where the black hole is in this particular coherent state.

The Coherent state as constructed in \cite{hall, tt} is peaked at the classical value
of the canonical pair $h_e(A), P_e(A, E)$ by construction. However by the nature of the
peak, the expectation values of the momentum take continuum values only for $P_e>>1$.
So for $P_e\sim t$ where $t=l_p^2/a_N$ ($l_p$ is Planck length, $a_N$ is an undetermined constant which is not present in the final result) is the semiclassicality parameter, one gets the expectation values
to be discrete. One could interpret this as a failure of the Coherent State to sample classical
values, or {\it one could say that the black hole is in a Coherent State}, in which case the spectrum
would be discrete. So just like `Light propagation' is produced from `Coherent States' in Electro Magnetism,
we can assume that the Classical Schwarzschild space-time is produced from these Coherent States, except
for very small values of the momenta, where the space-time is reproduced from the following spectrum \cite{adg}.
\be
P_{e_r} = (j + \frac12)t
\label{quant}
\ee
The above is the gauge invariant momentum for the radial edge. And explicitly the classical $P_{e_r}$ can be computed on the dual surface which is a piece of $S^2$; it has the following form:
\begin{widetext}
\bea
P_{e_r} =
\frac{r_g^2}{a_N~\a^4}\left[ \left(\frac{\sin[(1-\a')\th]}{(1-\a')}\right)^2  + \left(\frac{\sin[(1+\a')\th]}{(1+\a')}\right)^2  - 2\cos(2\th_0)\frac{\sin[(1-\a')\th]}{1-\a'}\frac{\sin[(1+\a')\th]}{1+\a'}\right]^{1/2} \label{mom}
\eea
\end{widetext}
With, $r_g$ the position of the classical horizon, $r$ the radial distance from the centre of the sphere, $\a'= \sqrt{\frac{1+\a^2}{2}}$, $\a=\sqrt{\frac{r_g}{r}}$, $\th_0$ the angle at which the radial edge intersects the 2-surfaces which constitute the `dual sphere', and $2\th$ the angular width of the dual surface. 
Writing the above in a more convenient form 
with the approximation $(1 \pm \a')\th <<1$:
\be
P_{e_r}= \frac{2 r^2}{a_N} \sin\th_0 \th \label{are1}
\ee
which one can see approximates the area for the two surface $S$ whose coordinate limits are $\th_0 -\th$ to $\th_0 + \th$ (In units of $a_N$). There is also a contribution from the $\phi$ coordinate, which is of the form $\e$, as expected is just the width of the
surface along the $\phi$ direction \cite{adg}. Now, there is an interesting aspect to note, and that is the fact that whenever $P$
becomes oscillatory as a function of $\a$ as observed in \cite{adg}, one finds that the above approximation fails!
This gives a restriction on the size of the graph and the graph degree of freedom to approximate the geometric
quantities like the triad. Another crucial point to note is the following, when one approximates the classical momentum
thus, ($(1 \pm \a)\th \ll 1$) all the other components of the momenta go to zero, except $P_1$ \cite{adg} (Note: the $\pm$ sign in the terms in $X_1$ and $X_3$ as given in \cite{adg} are typo errors and should be reversed), which is a special direction in the internal SU(2) space. This is a signal for the break down of SU(2)
degrees of freedom to U(1), a similar property noticed previously in \cite{cant}. Thus on any spherical 2-surface, the SU(2) degrees of
freedom must break down to U(1) to agree with classical physics. 

But what is special about the Horizon?
From (\ref{mom}), the classical horizon is at $\a\rightarrow 1$. 
Clearly for $\a=1$, there is a special limit, and as $\a\rightarrow 1$ $P_{e_r}$ tends to the following
\bea
P_H &= &\frac{r_g^2}{a_N} \left[ \th^2 + \frac{\sin ^2 2\th}{4}\right. \nn\\ &&\left.- 2\cos 2\th_0 \th \frac{\sin 2\th}{2}\right]^{1/2} \label{hor}  
\eea
Thus again, for small size of the width of the graph along $\phi$ direction,
this is indeed the classical area of the two dimensional `bit' (\ref{are1}) induced by the radial edge which intersects the 2-surface at angle $\th_0$ at $r=r_g$.
By (\ref{quant}) the radial edges can lie only at discrete values of $\th_0$,and the
area which is measured by $P_{e_{ri}}$  associated with the ith edge is $2 r_g^2\e~\th^i\sin\th_0^i$.
A sum of these bits should eventually give the total horizon. Also, since in the classical
variables the Immirzi parameter $\b=1$ corresponds to the Schwarzschild solution, we retain the same.

Now, to understand why $P_{e_r}$ measures the area of the spheres, one examines 
the definition of area of any infinitesimal 2-surface. As we show below, the radial edge which intersects the horizon induces a area \cite{thiem1}:
\be
A_{e} =  \sqrt{E_I (S) E_I(S)}
\ee
Where, given surface $S$ is infinitesimal and $E_I (S)=\int _S* E_I$.
Now, 
\bea
P_e^I = -\frac1{a_N}Tr[T^I h_e\int_S h_{\r}* E h_{\r}^{-1} h_e^{-1}] 
\eea

With $h_e$ being the holonomy of the edge, and $h_{\r}$ being holonomies on the
edges lying on $S$. For sufficiently small $S$, $P_{e_r}= \sqrt{ P^I_e P^I_e} \rightarrow
\sqrt{E_I(S) E_I (S)} \sim A_e$. This is why we have (\ref{are1}) giving you the area at the classical level. Thus, it is needless to
say that for small bit of the Horizon, the area induced by the ith radial edge must be (The $a_N$ cancels)
\be
A_H^i= (j^i+ \frac 12) l_p^2
\ee
This surprisingly is the spectrum proposed in \cite{poly} as the correct
one for Horizon area. This confirms the equidistant nature of the spectrum
and as shown in \cite{poly},one can assume a thermal bath, and derive 
Hawking temperature etc. Here we derive the microscopic entropy associated 
with the above Horizon picture.
The degeneracy associated with ith bit of the Horizon associated with the ith radial edge is:
\be
(2 j_i + 1)^2
\ee
(Recall the coherent state in the momentum representation is $ e^{-t(j(j+1))}\pi_{j mn}$, where j is the eigenvalue
of the casimir, and m and n are the quantum numbers running from $-j,..., j$ and which produce
the above degeneracy \cite{tow}.) There is a further restriction to only $m$, as the edges can only be ingoing
at the Black Hole horizon. From all previous derivations, one point is definitely different, and that is the geometric interpretation of the spectrum. Since here we can assume a direct relation to the geometry of the space-time, and the details
of the graph (width :$\th\e$, intersection point $\th_0$), one can conjecture on the nature of the Spins. Since $\sum A_{H}^i = A_H$, 
$j_i$ being arbitrary would imply that the number of radial edges crossing the horizon would be
asymmetric distributed as $\th_0^j$ would not be evenly spaced. The most symmetrical configuration with the
maximum number of radial edges is the situation where the partition of $A_H/l_p^2$ is done by 1. This corresponds
to $j_i =1/2$. The number of radial edges is then $N= A_H/l_p^2$. The degeneracy is:
\be
\prod_i^N (2) = 2^{N}
\ee
The Entropy in this case is 
\be
S_{BH} = \frac{A_{H}}{l_p^2}~\log 2
\ee
Now, the Immirzi parameter which fixes the spectrum of $P$ can be $4 \log 2$. The Immirzi parameter appears as the semiclassicality parameter $t$ picks up $\b$ in the different quantisation sectors of the theory \cite{tt}, whereas the classical quantities remain the same. However, since the counting has been performed
in a rudimentary fashion, this value of $\b$ should not be taken seriously. 
Different predictions on that and more on Black Hole entropy from canonical quantum gravity 
can be found in \cite{cant,quasi}. However the present derivation gives the semiclassical physics a more genuine note by
linking the counting to expectation values of geometric operators in coherent states.

Now we come to the answers to the questions we had raised earlier.\\ 
1)Holography: To illustrate the point we are making, let us examine the Volume Operator:
This is given by \cite{sahl}:
\be
V \propto \sqrt{\sum_{\{e_1,e_2,e_3\}}\sum_{e,e',e''\subset \{e_1,e_2,e_3\}} \e_{e e' e''} \e_{IJK} P^I_eP_{e'}^J P_{e''}^K}
\ee
Where $e,e'e''$ constitute triplet of edges meeting at a particular Vertex. Obviously one of the edges
is radial. Since as observed in \cite{adg}, the expectation values
of $P_{e_\th}, P_{e_\phi}$ momenta take continuum values even for small black holes, and counting them is quite meaningless, the only degeneracy counting shall come from $P_{e_r}$. Thus the microscopic degrees of freedom for a given classical configuration, shall again be those associated with $P_{e_r}$, even though one is measuring the volume of the system. This is a typical realisation of the nature of Holography of the black hole space-time.\\
2)Why is the Counting correct for the Horizon $S^2$? As stated in the above, the graph degrees
of freedom are sampled by $h_e(A), P_e(A,E)$, and their counting alone should give the degrees
of freedom associated with the black hole. Thus even though in any arbitrary $S^2$ the degeneracy associated with $P$ can be proportional to area, this degeneracy might be broken by any arbitrary function of $h_e$ whose expectation values depend on the quantum numbers $m,n$. It is only at the horizon $S^2$ , that the canonical pair are restricted by the trapped surface equation. [The assumption is that the quantum fluctuations preserve the $S^2$ or respect the spherical symmetry of the Horizon: A fluctuation in $h_{e_r}\propto m$ implies a offdiagonal $(K_{r\phi}\propto A_r^3E_{\phi}^3)$  extrinsic curvature]. And hence {\it only at the horizon}, 
for any operator measurements, degrees of freedom $\equiv$ degeneracy of $P_H$.\\
3)Why is the degeneracy associated with the Horizon the entropy of the Black Hole?
As written down in \cite{adg}, the Coherent State Wave function extends into the black hole. However,
the constant time slice chosen is that of the proper-time observer, who falls into the black hole
and for her/him, the horizon does not play any special role. But, these coordinates fail for the
asymptotic observer, and one must resort to the Schwarzschild time to get the correct frame. In these
the Horizon is the latest two surface one can observe as $r-r_g\propto e^{-t/r_g}$. The Coherent State inside
the Black Hole must be traced over to produce a density matrix: $\rho = \sum_j |\psi_O>|\psi_H>|\psi_I^j><\psi_I^j|<\psi_H|<\psi_O|$. The expectation values yield the classical results, hence once one has traced
over the 'inside' coherent state, one is left with the Horizon and the Outside. 
As per the graph we have chosen: with radial spokes, fixing the $j_i$ at the horizon
fixes all subsequent $j_i$ at different radii outside the Horizon. Thus, a given
list of $j_i$ at the horizon is like fixing a boundary condition, and state
through out the outside is determined. In other words, the degeneracy associated with the Horizon Area
or the operator $P_H$ determines the complete number of microscopic degrees of freedom 
for a given Black Hole Coherent State. Hence modulo the Immirzi parameter, indeed, the black hole
`Bekenstein-Hawking' Entropy is
$$S_{BH} = \frac{A_H}{4l_p^2}.$$

\noindent
{\bf Acknowledgement:} I am grateful to S. Das for comments on the manuscript and T. Thiemann for useful correspondence.
This Work is supported in part by the ``Actions
de Recherche Concert{'e}es" of the ``Direction de
la Recherche Scientifique - Communaut{'e} Francaise
de Belgique", by a ``P{o}le d'Attraction Interuniversitaire"
(Belgium), by IISN-Belgium (convention 4.4505.86) and
by the European Commission RTN programme HPRN-CT-00131,
in which ADG is associated to K. U. Leuven.

\end{document}